\documentclass[aps,pra,english,groupedaddress, 
twocolumn, 10pt]{revtex4-2}
\usepackage{graphicx}
\usepackage{amsmath}
\usepackage{amssymb}
\usepackage[font=small,labelfont=bf]{caption}
\usepackage{amsfonts}
\usepackage[margin=2cm]{geometry}
\usepackage{mathtools}
\usepackage{multirow}
\usepackage{array} % Include this in your preamble for better table styling
\usepackage{subcaption}
\usepackage[labelfont=bf, labelsep=period, font=small]{caption}
\usepackage{dcolumn}
\usepackage{seqsplit}
\usepackage{relsize}
\usepackage{mathrsfs}
\usepackage{calrsfs}
\usepackage{booktabs}
\usepackage{caption}
\usepackage[usenames,dvipsnames]{color}
\usepackage{hyperref}
\hypersetup{
     linkcolor=blue,        % internal links (sections, equations)
    citecolor=teal,        % citations
     filecolor=magenta,     % file links
     urlcolor=violet,       % \href and \url links
}

\begin{document}
%\pagecolor{blue!40}
%\color{black}
%
\title{Endoreversible Carnot machines
with long-time reversible limit}
\author{Ramandeep S. Johal} 
	\email[e-mail: ]{rsjohal@iisermohali.ac.in}
\author{Rajeshree Chakraborty}
	\email[e-mail: ]{mp18011@iisermohali.ac.in}
	\affiliation{Department of Physical Sciences,\\
		Indian Institute of Science Education and Research Mohali,
		Sector 81, S.A.S. Nagar,\\
		Manauli PO 140306, Punjab, India.}
\begin{abstract}
The well-known Curzon-Ahlborn (CA) engine 
[Am. J. Phys. {\bf 43}, 22 (1975)] 
runs in finite time and yields a finite power output. 
In order to recover the standard Carnot cycle from the CA model, 
the limit of quasi-static operation
has to be concurrent with 
the reversible limit, a condition strangely missing from
the CA model. We augment this model by specifying
the duration of the isothermal process as 
a monotonic decreasing function of the temperature
difference between the working substance and
the reservoir.  As an illustration, 
we analyze the machine's operation in 
the low-dissipation (LD) regime in which the entropy production
in each isothermal branch is inversely proportional to 
the duration of the process. The modified model,  while retaining the optimal
 features of the CA model, also yields 
 explicit expressions for 
the optimal durations of isothermal processes. 
{We compare the optimal cycle periods of 
the present model and the LD model, and
show their agreement in the linear response
regime.}
\end{abstract}
\maketitle
{\it Introduction}: Despite the fundamental significance 
of Carnot cycle in thermodynamics  \cite{Carnot1824}, 
it is unrealizable since
the isothermal branches 
proceed infinitely slowly. 
To speed up the cycle,
the working substance (henceforth addressed as the system) and a heat reservoir must not be in mutual thermal equilibrium, 
thus allowing the transfer of heat at a finite rate. 
The simplest, though quite idealized, model of
a finite-time thermal machine is the 
endoreversible model 
\cite{CA1975, Rubin1979a, Hoffman2008}, 
in which the system undergoes reversible transformations 
while any irreversibility 
lies in the interaction between the system and the environment.
The broader area of research
is known as 
 finite-time thermodynamics in the Physics community \cite{Andresen1984},   
or as entropy generation minimization in the Engineering parlance
\cite{Bejanbook1995}. 
Assuming 
Newton's law of heat transfer,
Curzon and Ahlborn (CA) optimized
the power output of the endoreversible
engine \cite{CA1975} and showed that the efficiency
at maximum power is given by 
an elegant formula: 
$\eta_{\rm CA}^{} =1-\sqrt{T_c/T_h}$.
The actual form of the efficiency 
depends on the choice of the heat transfer 
law \cite{Chen1989}.
Despite the 
idealizations involved, the CA formula 
is often used to benchmark the efficiency of 
actual plants or engines. The model 
has motivated  new research directions 
and a vast literature 
\cite{Andresen2022, Andresen2025half}. 

Despite the simplicity and popularity of
the CA model, it shows some peculiarities that,  
to the best of our knowledge,
have not been addressed before. 
The control variables of the model are 
the two temperatures ($T_1$, $T_2$)
of the system and the two durations 
($t_h$, $t_c$) of the isothermal processes. 
However, the maximum power solution determines  
only the ratio of isothermal durations \cite{CA1975}, 
while leaving their absolute values
undetermined.
As a consequence, the total cycle 
time as well as the absolute magnitudes
of the heat and work at the optimal cycle are
also indeterminate. 
The apparent reason for this underspecification  
is the lack of as many 
equations in the CA model
to determine all the variables. 
As the power expression  
depends only upon two of the variables, for example
$T_1$ and $T_2$, 
its optimization over these does not pin down the two durations, 
but rather only their ratio \cite{Comment_time}. 

Further, one expects the CA model to recover
the standard Carnot cycle, implying that the 
reversible limit be concurrent with the quasi-static limit. 
Now, to have a reversible heat exchange,
the temperature 
difference between the system and a reservoir during each isothermal process must be infinitesimal. But, a finite
amount of heat is transferred only if the duration of the process 
diverges as the temperature difference vanishes.
Thus, the limits of a vanishing temperature
difference and of a quasi-static operation 
have to be applied together---in order to 
recover a reversible cycle.

On the other hand,  simply applying 
the quasi-static
limit at a finite temperature difference (between system and reservoir), 
does not yield reversible operation. 
Thus, we observe that 
the control over the isothermal process duration is distinct from the control
over the temperature difference.
While this does not hinder the determination 
of the maximum power and the efficiency thereof,
other relevant quantities---which depend explicitly on the process durations, 
remain undetermined. 
In this letter, we augment
the CA model and close the above gap by specifying a definite
relation between  
these two controls such that they may be regarded as equivalent. By construction,
the modified model approaches quasi-static limit  
when we apply the reversible limit, and vice versa. 
The model is then able to specify the absolute 
magnitudes for all the time scales, and by 
extension, the energy scales involved in the cycle.  
As will become clear, the additional premise  
makes the CA model 
compatible with another 
well-studied finite-time model
of Carnot engines with reversible limit, known as   
the low-dissipation (LD) model \cite{Esposito2010}.

We first formulate a generic  
relation between the isothermal duration  
and the temperature difference between the reservoir and system that ensures a 
a reversible operation in the quasi-static limit. Consider the CA model 
as depicted in Fig. 1. 
We first analyze the 
situation at the hot reservoir, where
$\delta T_1 = T_h -T_1 >0$. 
Now, the process of heat exchange is assumed to take 
sufficiently long so that, for a given gradient $\delta T_1$, 
a steady-state heat flux can be assumed.  
Also, the heat flux ($\dot{Q}_h$) passing
through the heat exchanger ($K_h$) is a monotonically
increasing function of the temperature 
difference: $\dot{Q}_h = F(\delta T_1)$, with $F(0)=0$, where the function $F(\cdot)$  specifies the heat-transfer law.
So, the heat transferred in time interval 
$t_h$ is just given by, $Q_h = F(\delta T_1) t_h$.

{The irreversible entropy produced is then given by 
\begin{align}
 S^{(\rm ir)}_{h} & = Q_h \left(\frac{1}{T_1} -\frac{1}{T_h}  \right), \nonumber \\
 & = \frac{F(\delta T_1)\delta T_1}{T_h(T_h-\delta T_1)}t_h
  \equiv {\cal F}_{T_h}^{}(\delta T_1)t_h.
  \label{dsirf}
\end{align}
Here, we have defined a function ${\cal F}_{T_h}^{}(\delta T_1)$,
for a given reservoir temperature $T_h$.
Clearly, 
${\cal F}_{T_h}^{} (\delta T_1)$ is monotonically increasing ($\delta T_1 >0$),
with  ${\cal F}_{T_h}^{} (0) =0$.
For instance,  $\delta T_1 \to 0$ implies a  
reversible operation ($S^{(\rm ir)}_{h} \to 0$).} Then, in order to exchange a
finite amount of heat (as in the standard 
Carnot cycle), the process must 
take infinitely long. 
We are interested in isothermal 
processes that 
not only operate with large durations, 
but also approach the reversible limit as 
$t_h \to \infty$ (equivalently, 
$x_h = 1/t_h \to 0$).
So, apart from the assumption of
a specific heat-transfer law, we assume that 
the entropy production is a monotonically
increasing function of $x_h$, i.e.
$S^{({\rm ir})}_{h} = G(x_h)$, with $G(0)=0$  denoting the reversible operation 
in the quasi-static limit.

Equating the above form of entropy production
with Eq. (\ref{dsirf}), and after rearranging,
we can write
${\cal F}_{T_h}^{}(\delta T_1) = x_h G(x_h) \equiv 
{\cal G}(x_h)$, where 
${\cal G}(\cdot)$ is also an increasing function.
Using the inverse of ${\cal G}^{}$, we can 
write $x_h = ({\cal G}^{-1}\circ {\cal F}_{T_h}^{})(\delta T_1)$,
or the duration of
the hot isothermal process, with a 
long-time reversible limit, may be written as   
\begin{equation}
    t_h = [({\cal G}^{-1}\circ {\cal F}_{T_h}^{})(\delta T_1)]^{-1}. 
    \label{genth}
\end{equation}
The monotonic increasing
property of the function ${\cal G}^{-1}\circ {\cal F}_{T_h}^{}$ implies that $t_h$ is monotonic decreasing
function of $\delta T_1$. More precisely,
 we obtain $t_h \to \infty$ 
as   $\delta T_1\to 0$. 
Using this  relation,
we can  
eliminate $t_h$ in favor of $\delta T_1$ 
and write, 
$Q_h = F(\delta T_1)/({\cal G}^{-1}\circ {\cal F}_{T_h}^{})(\delta T_1)$, effectively reducing the 
number of variables by one. 
A similar treatment on the side of the cold contact 
yields a definite one-to-one relation 
between the time interval of the cold
isothermal process  $t_c$ and $\delta T_2 = 
T_2-T_c$ (see Fig. 1). In the following, we make a specific choice of the (monotonic increasing) 
functions $F$ and $G$ to 
analyze the performance of the 
finite-time Carnot engine.
\begin{figure}
    \centering
    \includegraphics[width=0.3\linewidth]{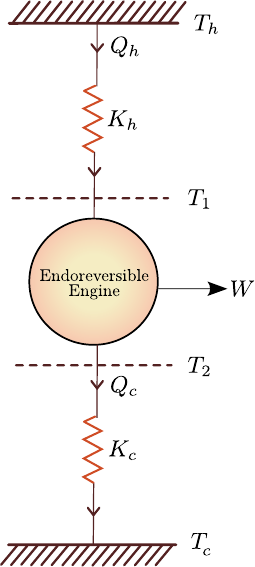}
    \caption{An endoreversible engine works between two temperatures $T_1 > T_2$ while
    the heat exchange with the hot ($T_h$)
    and cold ($T_c$) reservoirs occurs 
    through a heat exchanger of thermal conductance $K_h$ and  $K_c$, respectively.}
    \label{fig1}
\end{figure}
\par\noindent 
{\it Newton's law}:
As the simplest case, we take both $F$ and $G$
as {\it linear} functions. 
The choice, $F(\delta T_1) = K_h 
\delta T_1$, with $K_h$ as the thermal 
conductance of the heat exchanger 
on the hot-end side, defines the Newton's law, employed
in the CA model \cite{CA1975}. 
The heat absorbed by the system is then 
given by 
\begin{equation}
{Q}_{h}  =K_{h}\left(T_{h}-T_{1}\right) t_{h}.
\label{Qh}
\end{equation}
 Secondly, the  choice of the function $G(x_h) \propto x_h$ 
is popularly known as 
the low-dissipation (LD) regime \cite{Esposito2010, Broeck2013,
Ma_PRL2020}.
For concreteness, in this letter, we follow 
Ref. \cite{Esposito2010} so that 
$G(x_h) = \Sigma_h x_h = \Sigma_h /t_h$, where $\Sigma_h >0$ is a prespecified, phenomenological
dissipation constant. 
Eq. (\ref{dsirf}) is then expressed in two equivalent forms:
\begin{equation}
    S^{(\rm ir)}_{h} = \frac{K_h(T_h-T_1)^2}{T_h T_1}t_h 
     = \frac{\Sigma_h}{t_h}.
     \label{dsirfN}
\end{equation}
From the above, we obtain 
\begin{equation}
t_{h}=\sqrt{\frac{\Sigma_{h} }{K_{h}}} 
\frac{\sqrt{T_{h} T_{1}}}{(T_{h}-T_{1})}.
\label{th}
\end{equation}
This gives an exact dependence of
$t_h$ on $(T_h-T_1)$, ensuring that 
the LD assumption 
is followed. 
Eliminating $t_h$ from Eq. (\ref{Qh}),
we can write 
\begin{equation}
Q_{h}=\sqrt{\Sigma_{h} K_{h} T_{h} T_{1}}.
\label{Qhh}
\end{equation}
A similar analysis is performed  
 for the heat rejected
 to the cold reservoir, given by  
 ${Q}_{c}  =K_{c}\left(T_{2}-T_{c}\right) t_{c}$,
 where $T_2 (> T_c)$ is the fixed temperature
 of the system, with duration of the isothermal process as $t_c$ and 
 thermal conductance $K_c$ (see Fig. 1). 
 Again assuming the LD regime, implying  
 $\Delta_{\rm i r} S_{c} ={\Sigma_{c}}/{t_{c}}$, 
 we obtain the expressions
\begin{equation}
 t_{c}=\sqrt{\frac{\Sigma_{c}}{K_{c}}} 
 \frac{\sqrt{ T_{2} T_{c}}}{(T_{2}-T_{c})},
 \label{tc}
 \end{equation}
and 
\begin{equation}
    Q_{c}=\sqrt{\Sigma_{c} K_{c} T_{c} T_{2}}.
    \label{Qcc}
\end{equation}
Now, the   
endoreversible condition  
stipulates  \cite{CA1975} 
\begin{equation}
 \frac{Q_h}{T_1}=\frac{Q_c}{T_2}, 
 \label{endorev}
\end{equation}
implying that there is no entropy production 
in the work generating compartment.
Using Eqs. (\ref{Qhh}) and (\ref{Qcc}), 
this condition yields 
\begin{equation}
T_{2}= \kappa \sigma \theta T_{1}, \label{T2andT1}
\end{equation}
where $\kappa={K_{c}}/{K_{h}}$,  
$\sigma={\Sigma_{c}}/{\Sigma_{h}}$,  
and $\theta={T_{c}}/{T_{h}}$. We also 
define the reduced temperature 
$\tau_{1}^{} = T_1/T_h <1$ and $\tau_{2}^{}
= T_2/T_h <1$.
The design condition  
$0< \tau_{2}^{} < \tau_{1}^{}$ implies 
that $\kappa \sigma \theta < 1$. 

The work output per cycle, $W= Q_h-Q_c > 0$, is given by
\begin{align}
W(\tau_{1}^{}) &=T_{h} \sqrt{\Sigma_{h} K_{h}} 
(1- \kappa \sigma \theta)\sqrt{\tau_{1}}, \nonumber \\
       & = (1- \kappa \sigma \theta)Q_h,
\end{align}
so that the efficiency, $\eta = W/Q_h$, 
simply is 
\begin{equation}
\eta=1- \kappa \sigma \theta.
\label{eff}
\end{equation}
Interestingly,  the efficiency 
of this model is fixed, depending  only
on the given parameter ratios $\kappa$, $\sigma$ and $\theta$. 
To comply with the Carnot bound 
 ($\eta <\eta_{\rm C}^{} = 1-\theta$),
we expect the condition $\kappa \sigma > 1$ for
irreversible operation. On the other hand, 
the reversible limit ($\kappa \sigma \to 1$) can be
understood as follows.
The vanishing of entropy production at both 
the hot and the cold contacts requires 
$T_1 \to T_h$ and $T_2 \to T_c$, respectively. 
In this limit, we approach the Carnot cycle,
whereby Eqs. (\ref{Qhh}) and 
(\ref{Qcc}) simplify to 
${Q}_{h} \to T_h \sqrt{\Sigma_{h} K_{h}} \equiv Q_{h}^{(0)} $
and ${Q}_{c} \to T_c \sqrt{\Sigma_{c} K_{c}} \equiv Q_{c}^{(0)}$.
The reversibility condition, 
${Q}_{h}^{(0)}/ T_h = {Q}_{c}^{(0)}/ T_c$, 
then yields $\Sigma_{h} K_{h} = \Sigma_{c} K_{c}$, or, $\kappa \sigma =1$. 

\par\noindent 
{\it Optimization of the power output}:
{To formulate the simplest of the finite-time heat cycles,
we assume that the durations of the adiabatic branches can 
be neglected in comparison to the isothermal branches \cite{Callenbook}.
This becomes feasible if we assume a working substance
with a short enough relaxation time so that the
expansion and compression processes can be performed 
in a fast manner.} 
Thereby, the total cycle time may be regarded as
$t= t_h + t_c$, and given by
\begin{equation}
t=\sqrt{\frac{\Sigma_{h} }{K_{h}}}\left(\frac{1}{1-\tau_{1}}+\frac{\sigma}{ \kappa \sigma \tau_{1}-1}\right) 
\sqrt{\tau_{1}^{}}, 
\label{tt}
\end{equation}
where we used Eqs. (\ref{th}), (\ref{tc}) and 
(\ref{T2andT1}).
The power output per cycle,  
$P(\tau_{1}^{}) = W/t$, is a function only 
of one variable $\tau_1$, for given values of the 
system/process parameters $K_i,\Sigma_i, T_i$, with $i=h,c$. 
Optimizing the power, 
we obtain the hot-side optimal temperature of the system: 
\begin{equation}
\tau_{1}^{*}=\frac{1+\sqrt{\kappa} \sigma}{(1+\sqrt{\kappa}) \sqrt{\kappa} \sigma}.
\label{tau1st}
\end{equation}
The optimal power reads as 
\begin{equation}
P\left(\tau_{1}^{*}\right)= K_{h} T_{h} \frac{( \kappa \sigma - 1)(1- \kappa \sigma \theta)}{\sigma(1+\sqrt{\kappa})^{2}}.
\label{optp}
\end{equation}
From Eqs. (\ref{th}), (\ref{tc})
and the relation $\tau_2 = \kappa \sigma \theta \tau_1$, 
we can explicitly determine the two durations. 
Their ratio depends only on the parameter $\kappa$ as 
\begin{equation}
    \left.\frac{t_{c}}{t_{h}}\right\vert_{\tau_{1}^{*}}=\frac{1}{\sqrt{\kappa}}. 
\end{equation}
From Eqs. (\ref{eff}) and (\ref{optp}), we can notice  
the power-efficiency trade off. As 
$\kappa \sigma \to 1$, the efficiency approaches
the Carnot bound, but the power output vanishes. On the other hand, for  $\kappa \sigma \to 1/\theta$, both the 
efficiency and the power approach zero. In other words,
this points to a parabolic curve between efficiency
and power, with the power attaining a maximum
for some intermediate value of efficiency. 
This is seen more clearly if we 
recast Eq. (\ref{optp}) in terms of efficiency. 
The optimal power at the given efficiency 
$\eta$ is given as 
\begin{equation}
P^*(\eta)=\frac{K_{c} T_{h}}
{(1+\sqrt{\kappa})^{2}} 
\frac{(\eta_{\rm C}^{} - \eta)\eta}{1-\eta},
\label{optpf}
\end{equation}
which shows that 
$P^*$ is maximized when the efficiency
is set at $\eta = \eta_{\rm CA}^{}$. Alternately, 
for a given configuration with specified values of 
$K_h, K_c, T_h$ and  $T_c$, 
 we obtain the maximum power 
when we set 
 $\sigma = 1/\kappa \sqrt{\theta}$,
 at which 
 the efficiency takes up the value 
 $\eta_{\rm CA}^{}$. 
The doubly optimized or the maximum power is given by:
\begin{equation}
    P_{\rm max} = 
    \frac{K_h K_c}{(\sqrt{K_h} + \sqrt{K_c})^2}
    \left( \sqrt{T_h} - \sqrt{T_c}\right)^2,
\end{equation}
which matches with the maximum power of
the CA model \cite{CA1975}.   Furthermore, 
the optimum hot-side temperature
of the system is also given by:
    $T_{1}^{*} =  (T_h + \sqrt{kT_hT_c})/(1+\sqrt{k})$. 
We emphasize that
the maximum-power
point of the CA model is retained in the present 
approach. 
After obtaining the optimal values
of $T_1$ and $T_2$,
we can determine
the corresponding durations from Eqs. (\ref{th}) and (\ref{tc}), and so 
the present model  computes all 
the relevant quantities at maximum power,
which was not possible in the original 
model. 

It is instructive to 
consider the entropy production 
in terms of the entropy changes of the system and
the reservoirs, as  
\begin{align}
S^{(\rm ir)}_{h} & = \Delta S - \frac{Q_h}{T_h}, \\
S^{(\rm ir)}_{c} & =  \frac{Q_c}{T_c}-\Delta S,
\end{align}
{where $\Delta S$ is the entropy increase (decrease) of the system
during the hot (cold) isothermal branch.}  
Now, the condition (\ref{endorev}) implies  
$\Delta S = Q_h/T_1 = Q_c/T_2$. Using the expressions
(\ref{Qhh}) and (\ref{Qcc}),
 we obtain  
\begin{equation}
    \Delta S = \sqrt{\frac{\Sigma_h K_h T_h}{T_1}} = \sqrt{\frac{\Sigma_c K_c T_c}{T_2}}, 
    \label{}
\end{equation}
showing the equivalence of the variables 
$T_1$ or $T_2$
and $\Delta S$. Using the above relations, 
we can write 
\begin{align}
S^{(\rm ir)}_{h} & = \Delta S - 
\frac{\Sigma_h K_h }{\Delta S} \geq 0, \label{dhds} \\
S^{(\rm ir)}_{c} & = 
\frac{\Sigma_c K_c}{\Delta S}
- \Delta S \geq 0. \label{dcds}
\end{align}
The positivity of each entropy-production
term implies the double inequality:
\begin{equation}
    \sqrt{\Sigma_c K_c} \geq \Delta S \geq 
 \sqrt{\Sigma_h K_h}.
\end{equation}
An equality in the above implies that  
the corresponding isothermal branch is
reversible. For a reversible cycle, 
both inequalities reduce to equalities,
yielding 
\begin{equation}
    \Delta S^{(0)} = \sqrt{\Sigma_{h} K_{h}} = \sqrt{\Sigma_{c} K_{c}}.
    \label{revs0}
\end{equation}
Note that the above relations 
further qualify the condition for reversibility, 
$k \sigma =1$, as
discussed  below Eq. (\ref{eff}).
The heat exchanges 
for the reversible cycle are then 
expressed in the standard form  
 $Q_{h}^{(0)} = T_h \Delta S^{(0)}$ and
$Q_{c}^{(0)} = T_c \Delta S^{(0)}$. 

{Clearly, for the CA model, the change in system entropy ($\Delta S$) deviates from the reversible value ($\Delta S^{(0)}$).
This is especially in contrast with the LD model  \cite{Esposito2010}
where the change in system entropy  
is constrained at $\Delta S^{(0)}$ as the irreversibility is
tuned. 
So, the two models differ in the applicable 
constraints. Further,
the CA model ascribes 
entropy production to heat flow
across a finite temperature gradient. 
The 
LD model rather simply assumes the form of entropy production
as an inverse dependence on the process duration,  
while leaving the
physical mechanism responsible for the entropy production unspecified. This makes the 
actual performance metrics of the 
CA model dependent on the 
choice of the heat transfer law, 
whereas the LD model seems to 
be based on generic assumptions.  
As to the specific  
predictions, the CA model is silent on the optimal durations of the isothermal
processes, whereas the LD model does 
provide explicit formulae for the same \cite{Esposito2010}. In this regard, our augmented CA model   
keeps the underlying physical picture of entropy 
production, while also incorporating the long-time 
reversible limit.
By using 
the LD assumption 
to formulate this limit, the CA model also yields    
explicit expressions for the durations of 
isothermal processes.}

{Thus, it is  
interesting to compare the process durations 
in the present model and the LD model. 
Substituting the maximum power conditions 
$\tau_1 = \tau_{1}^{*}$ [Eq. (\ref{tau1st})] and 
$\sigma =1/k\sqrt{\theta}$ into Eq. (\ref{tt}), we obtain the 
 period of the maximum-power CA cycle as
\begin{equation}
t_{\rm CA}^{} = \sqrt{\frac{\Sigma_h}{K_h}}
 \frac{(1+\sqrt{k})\sqrt{(1+\sqrt{k})(1+\sqrt{k\theta})}}{k(1-\sqrt{\theta})}.
 \label{tca}
\end{equation}
On the other hand, the corresponding period
for the maximum-power LD cycle is calculated to be 
\cite{Esposito2010}
\begin{equation}
t_{\rm LD}^{} = 
\frac{\Sigma_h}{\Delta S^{(0)}}
\frac{2(1+\sqrt{\sigma \theta})^2}{1-\theta}.
\label{tld}
\end{equation}
The prefactors $\sqrt{\Sigma_h/K_h}$ and  ${\Sigma_h} /{\Delta S^{(0)}}$ in the above
equations reflect the intrinsic time scales
in the respective models. 
To make a reasonable comparison between 
these equations, we choose
the same reversible limit for both,
i.e. each model reduces to the same reversible cycle with given values $T_h, T_c$ and $\Delta S^{(0)}$. Then, 
owing to Eq. (\ref{revs0}), 
we may identify the two time
scales, i.e.  
${\Sigma_h} /{\Delta S^{(0)}} = 
\sqrt{\Sigma_h/K_h}$. Together with the 
choice of an identical parameter value $\sigma = 1/k\sqrt{\theta}$,  we can rewrite Eq. (\ref{tld}) as 
\begin{equation}
t_{\rm LD}^{} = 
\sqrt{\frac{\Sigma_h}{K_h}}
\frac{2(\sqrt{k}+\sqrt{\sqrt{\theta}})^2}{k(1-\theta)}.
\label{tld2}
\end{equation}
The reciprocals of the expressions (\ref{tca}) and 
(\ref{tld2}) are compared 
in Fig. 2, for $k=1$. 
We observe an agreement in the near-equilibrium regime
($\theta \approx 1$), as corroborated
by the series expansion 
at small temperature differences, 
\begin{equation}
t_{\rm CA}^{-1} \approx t_{\rm LD}^{-1} 
= \sqrt{\frac{K_h}{\Sigma_h}} 
\frac{k(1-\theta)}{2(1+\sqrt{k})^2} + {\cal O}[(1-\theta)^2],
\label{sertcald}
\end{equation}
showing an agreement in the linear response regime,
for arbitrary $k$ values, with differences appearing
at the higher order terms. 
These findings are consistent with the earlier 
studies in which the optimal power performance
of LD and endoreversible models were found to be
comparable in the linear-response regime 
\cite{Johal2017, Iyyappan2019,  Zhang2020}. 
A detailed comparison of the two models based on other 
quantities, such as power output, efficiency and trade-offs 
between them \cite{Ma2018}, is beyond
the scope of this letter. 
We, however, emphasize that the LD assumption constitutes a simple analytic 
application of the present approach which,  in principle, extends to 
other models with a long-time reversible 
limit, but which may operate under far-from-reversible 
conditions \cite{Gerstenmaier2021}. }
\begin{figure}
\includegraphics[width=0.95\linewidth]{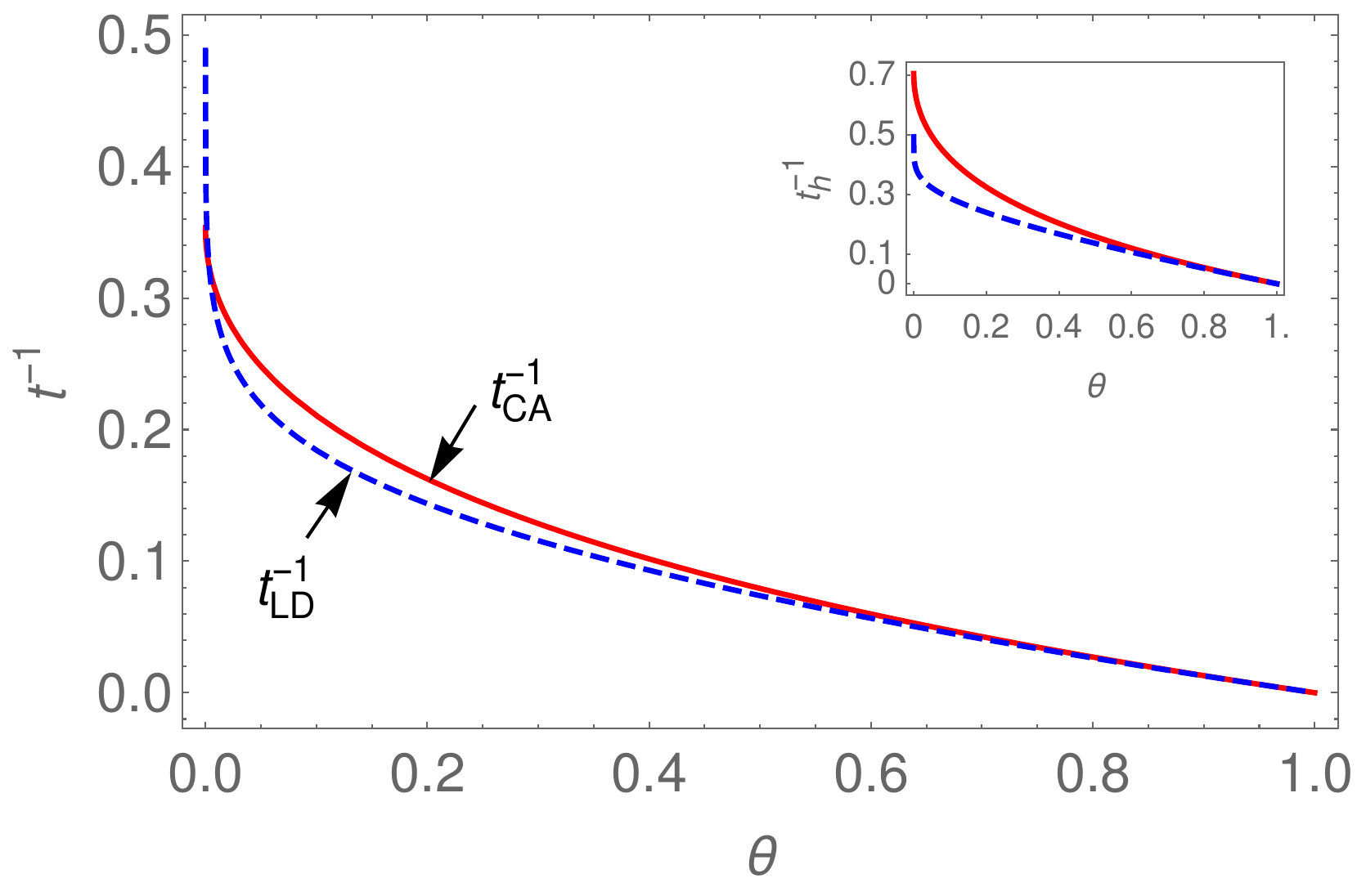}
    \caption{Comparison of reciprocal time periods at maximum power in the augmented CA model  and the LD model, using respectively Eqs. (\ref{tca}) and (\ref{tld2}) at $k=1$, each scaled by the factor $\sqrt{\Sigma_h/K_h}$. The inset shows the corresponding durations for the hot contact branches. }
    \label{fig:placeholder}
\end{figure}

Finally, the CA engine can  be treated exactly using other heat transfer laws,
such as the linear-irreversible law \cite{Chen1989} where the heat flux
is proportional to the difference of inverse 
temperatures. Upon incorporating the 
LD assumption on both hot and
cold isothermal branches, the heat 
exchanged with a reservoir is simply given by
$Q_{h} =\sqrt{\alpha_{h} \Sigma_{h}}$ 
and $Q_{c} =\sqrt{\alpha_{h} \Sigma_{c}}$.
Here, $\alpha_i$ is the heat conductivity
of the heat exchanger at the hot/cold contact. 
Therefore, the efficiency of the engine
is fixed at $\eta  = 1-\sqrt{\alpha \sigma}$,
where $\alpha = \alpha_c/\alpha_h$. 
The power is first optimized 
over the only variable $\tau_1$ (see Appendix A).
The power so obtained becomes maximum
when the given efficiency is 
\begin{equation}
\bar{\eta}=\frac{(1+\sqrt{\alpha})\eta_{\rm C}^{}}{2(1+\sqrt{\alpha})-\eta_{\rm C}^{}},    
\end{equation}
which matches with the known results \cite{Chen1989}.
As mentioned above, 
the modified model allows to calculate
explicitly the durations and other related quantities. 
Thus, from the above two heat transfer laws, we note that the optimal features
of the endoreversible model are preserved
under the inclusion of the LD assumption. 
A similar analysis can be performed
for the refrigerator mode, where 
the target function is chosen
as the product of coefficient of performance
and the cooling power (see Appendix B).

\par\noindent
{\it Conclusions}: 
The CA model at maximum power is unable to 
specify the optimal durations of isothermal processes.   
By augmenting the model  with a 
long-time reversible limit,  
each duration can be regarded 
equivalent to the corresponding control over the  
temperature difference between the system and reservoir. 
Thus, we are able to determine the durations 
as well as the absolute magnitudes of
the heat and work involved in the optimal cycle. 
The particular cases studied above show that 
the endoreversible model can be made
compatible with the low-dissipation assumption.
This additional assumption introduces 
a new scale in the CA model,
for example, the dissipation constant
of the LD model. 

An interesting consequence
of this extended model is a relation between the parameters 
belonging to different models, such as Eq. (\ref{revs0})
which relates the dissipation constant
to the heat transport coefficient, 
via the reversible entropy change of the reference
Carnot cycle. 
{We have also noted that the 
intrinsic time scales in the LD model 
and the present model are deemed equal
by referring to the same reversible cycle. 
This leads to the observation that 
the optimal process durations in the 
two models agree in the linear response
regime. 
It is hoped that
the augmented model not only provides a 
better logical description of
the endoreversible model, but can 
foster a deeper understanding of 
the relationship between different models. 
A corresponding analysis of more realistic models
beyond the LD
assumption \cite{Ma_PRL2020, Gerstenmaier2021}, or 
incorporating internal 
irreversibilities and heat leaks \cite{Calvo2017, Holubec2025},
are some of interesting future directions.} 

\section{End Matter}
\subsection{CA engine with linear-irreversible law}
In the following, we extend the CA engine
by adding a long-time reversible limit in the 
form of the LD assumption.
As per the linear-irreversible heat transfer law,
the heat flux is proportional to the 
difference of inverse temperatures,
across the heat exchanger of conductivity
$\alpha_i$ with $i=h,c$.
So, the heat absorbed from the hot reservoir is given by 
\begin{equation}
Q_{h}=\alpha_{h}\left(T_{1}^{-1}-T_{h}^{-1}\right) t_{h}.
\label{Qhlin}
\end{equation}
The irreversible entropy generated at the hot contact is
\begin{equation}
\begin{aligned}
S^{(\rm ir)}_{h} & =Q_{h}\left(T_{1}^{-1}-T_{h}^{-1}\right)  \\
& =\alpha_{h}\left(T_{1}^{-1}-T_{h}^{-1}\right)^{2} t_{h}. 
\end{aligned}
\end{equation}
Invoking the LD assumption, $S^{(\rm ir)}_{h}= {\Sigma_{h}}/{t_{h}}$, we obtain the hot-side contact time:
\begin{equation}
\begin{aligned}
t_{h} & = \sqrt{\frac{\Sigma_{h}}{\alpha_{h}}} {\left(T_{1}^{-1}-T_{h}^{-1}\right)}^{-1}.
\end{aligned}
\end{equation}
Substituting the above expression in Eq. (\ref{Qhlin}), we obtain 
\begin{equation}
Q_{h} =\sqrt{\alpha_{h} \Sigma_{h}}.
\end{equation}
Similar expressions can be derived for the cold contact, as
\begin{equation}
t_{c}=\sqrt{\frac{\Sigma_{c}}{\alpha_{c}}}\left(T_{c}^{-1}-T_{2}^{-1}\right)^{-1}, \qquad \quad Q_{c}=\sqrt{\alpha_{c} \Sigma_{c}}.
\end{equation}
Imposing the condition of endoreversibility, ${Q_{h}}/{T_{1}} = {Q_{c}}/{T_{2}}$, leads to
\begin{equation}
 T_{2} =\sqrt{\alpha \sigma} T_{1}.
\end{equation}
where $\sigma={\Sigma_{c}}/{\Sigma_{h}}$ and $\alpha={\alpha_{c}}/{\alpha_{h}}$.
The efficiency, $\eta = 1 -Q_c/Q_h,$ is given by
\begin{equation}
\eta  = 1-\sqrt{\alpha \sigma}.
\label{eff_lin}
\end{equation}
Thus, the efficiency is fixed and independent
of the reservoir temperatures. 
Consistency with the Carnot bound requires:
$\theta <  \sqrt{\alpha \sigma} <1$.
In terms of reduced temperatures,  $ \tau_{2}^{} =\sqrt{\alpha \sigma} \tau_{1}$. The work output is,  $W=Q_{h}-Q_{c} = \sqrt{\alpha_{h} \Sigma_{h}}\left(1-\sqrt{\alpha \sigma}\right)$,
while the total contact time is 
\begin{equation}
 t =T_{h}\sqrt{\frac{\Sigma_{h}}{\alpha_{h}}} {\left(\frac{1}{1-\tau_{1}^{}}+\frac{\sigma \theta}{\sqrt{\alpha \sigma} \tau_{1}^{}-\theta}\right)\tau_{1}^{}},
\end{equation} 
with $\theta=T_{c}/T_{h}$.
Thus the power output, $P= {W}/{t}$, reads
\begin{equation}
P\left(\tau_{1}\right)=\frac{\alpha_{h}(1-\sqrt{\alpha \sigma})}{ T_{h}\tau_{1}}\left(\frac{1}{1-\tau_{1}}+\frac{\sigma\theta}{\sqrt{\alpha \sigma}\tau_{1}^{}-\theta}\right)^{-1}.
 \end{equation}
Maximizing $P$ with respect to $\tau_{1}$ yields the optimal hot-end temperature:
\begin{equation}
\tau_{1}^{*}=\frac{(1+\sqrt{\sigma}) \theta}{(\sqrt{\alpha}+\theta) \sqrt{\sigma}}.
\end{equation}
The ratio of optimal contact times satisfies
\begin{equation}
\left(\frac{t_{c}}{t_{h}}\right)^{*}  = \sqrt{\sigma}.
\end{equation}
The corresponding maximum power is
\begin{equation}
 P\left(\tau_{1}^{*}\right)  =\frac{\alpha_{h}(\sqrt{\alpha \sigma}-\theta)(1-\sqrt{\alpha \sigma})}{T_{c}(1+\sqrt{\sigma})^{2}}.
\end{equation}
In terms of the efficiency, the above
power is written as
\begin{equation}
 P(\eta)  =\frac{\alpha_{c}}{T_c}
\frac{(\eta_{\rm C}^{} - \eta)\eta}{(1+\sqrt{\alpha} -\eta)^{2}}.
\label{peta}
\end{equation}
Over the class of engines having different 
efficiencies [Eq. (\ref{eff_lin})],
the power output is maximum 
at the efficiency:
\begin{equation}
\bar{\eta}=\frac{(1+\sqrt{\alpha})\eta_{\rm C}^{}}{2(1+\sqrt{\alpha})-\eta_{\rm C}^{}}.
\end{equation}
This efficiency is bounded as:
\begin{equation}
   \frac{\eta_{\rm C}^{}}{2} \leq \bar{\eta}
   \leq \frac{\eta_{\rm C}^{}}{2-\eta_{\rm C}^{}}.
\end{equation}
The upper and lower bounds are obtained
in the limit $\alpha \to 0$ and 
$\alpha \to \infty$, respectively. 
Finally, the maximum power at which 
we obtain the efficiency $\bar{\eta}$ is 
given by
\begin{equation}
    P_{\rm max}^{} = 
    \frac{\alpha_{c}}{4T_c}
\frac{\eta_{\rm C}^{2}}{(1+\sqrt{\alpha})
(1+\sqrt{\alpha} - \eta_{\rm C}^{})}.
\end{equation}
The above expressions are the exact results
of the endoreversible engine with
the linear-irreversible law \cite{Chen1989}.
Furthermore, we can calculate all the related
quantities such as $t_c$ and $t_h$, and so on.
In the original endoreversible model, it
was possible to evaluate only the ratios
such as $t_c/t_h$. 

%%%%%%%%%%%%%%%%%%%%%%%%%%%%%%%%%%%%%%%%%%%%%%%%%%%%
\subsection{Endoreversible Refrigerator with low-dissipation}
%%%%%%%%%%%%%%%%%%%%%%%%%%%%%%%%%%%%%%%%%%%%%%%%%%%%
The model of irreversible refrigerator is 
depicted in Fig. 2. Here, $T_c > T_2$ and 
$T_1 > T_h$. 
\begin{figure}
    \centering
    \includegraphics[width=0.5\linewidth]{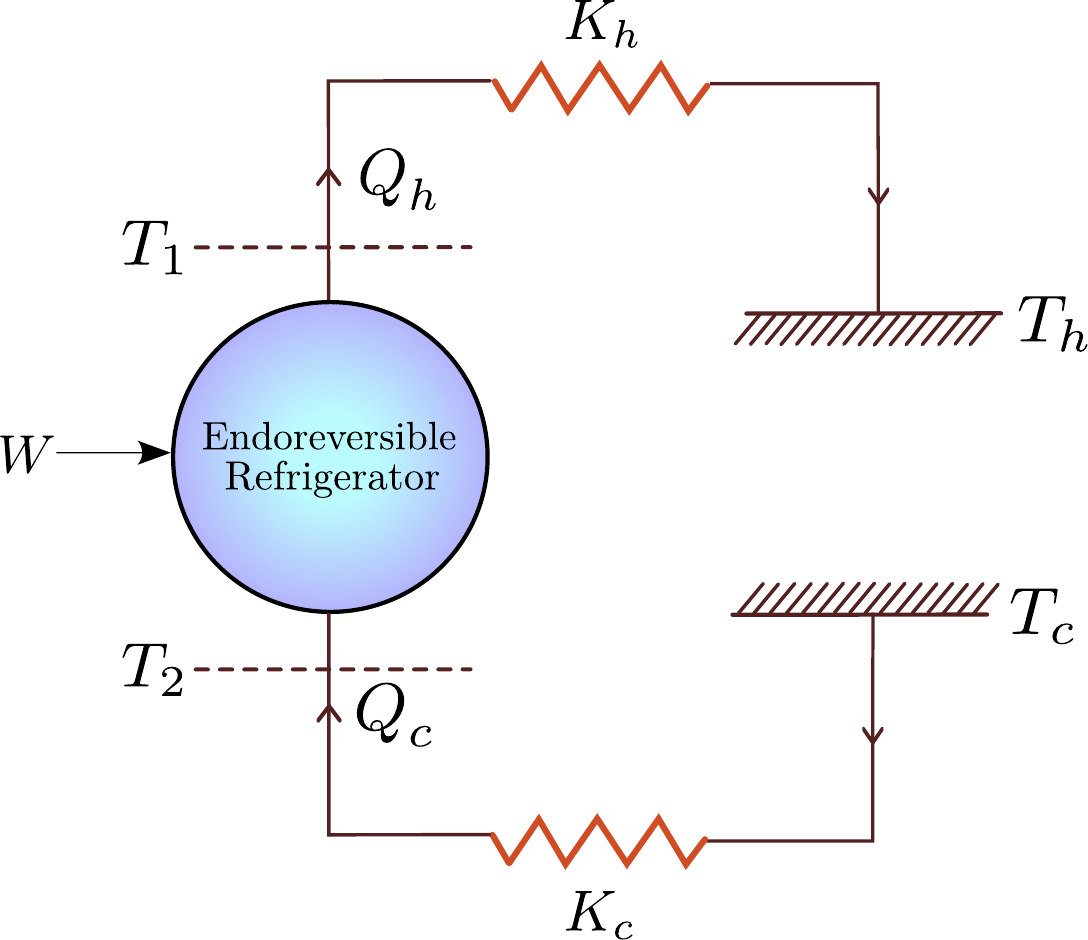}
    \caption{An endoreversible refrigerator operates between two fixed temperatures ($T_1 > T_2$)  
    by pulling heat $Q_c$ delivered from a 
    cold reservoir ($T_c > T_2$) via 
    heat exchanger of thermal conductance $K_c$ while
    dumping heat, $Q_h = Q_c +W$, to a hot reservoir 
    ($T_h < T_1$) via a heat exchanger
    of thermal conductance $K_h$. 
    The heat flow in heat exchangers 
    is assumed to follow Newton's law.}
    \label{fig:placeholder}
\end{figure}

Under Newton’s law, the heat absorbed from the cold reservoir is given by
\begin{equation}
{Q}_{c}  =K_{c}\left(T_{c}-T_{2}\right) t_{c} >0.
\label{qc}
\end{equation}
The associated entropy production, 
$S^{(\rm ir)}_{c}  = Q_{c}\left( T_{2}^{-1}-T_{c}^{-1} \right)$,
is
\begin{equation}
\begin{aligned}
S^{(\rm ir)}_{c} 
=\frac{K_{c}\left(T_{c}-T_{2}\right)^{2}t_c}{T_{c} T_{2}}. 
\end{aligned}
\end{equation}
Then, assuming the LD condition
$\Delta_{i r} S_{c} ={\Sigma_{c}}/{t_{c}}$, 
we can arrive at the following expressions:
\begin{equation}
 t_{c}=\sqrt{\frac{\Sigma_{c}}{K_{c}}} 
 \frac{\sqrt{T_{c} T_{2}}}{(T_{c}-T_{2})},  \qquad Q_{c}=\sqrt{\Sigma_{c} K_{c} T_{c} T_{2}}.
 \end{equation}
 Similarly, 
the hot-side contact time and the heat
rejected to the reservoir are obtained as
\begin{equation}
t_{h}=\sqrt{\frac{\Sigma_{h}}{K_{h}}} 
\frac{\sqrt{T_{h} T_{1}}}{(T_{1}-T_{h})}, \qquad 
Q_{h}=\sqrt{\Sigma_{h} K_{h} T_{h} T_{1}}.
\end{equation}
The endoreversibility condition yields the temperature relation $T_2 = \kappa \sigma \theta T_1$, which is the same as for the engine. Using the reduced definitions of temperatures, $\tau_{1} = T_1/T_h > 1$ and
$\tau_{2}^{} = T_2/T_h < 1$, we have  
$\tau_{2}^{}= \kappa \sigma \theta \tau_{1}^{}$. The work input is given by
\begin{align}
W &=T_{h} \sqrt{\Sigma_{h} K_{h}} 
(1- \kappa \sigma \theta)\sqrt{\tau_{1}^{}}, \\
  & = \frac{1- \kappa \sigma \theta}
  {\kappa \sigma \theta}Q_c.
\end{align}
The total cycle time is given by
\begin{equation}
t=\sqrt{\frac{\Sigma_{h} }{K_{h}}}\left(\frac{1}{\tau_{1}-1}+\frac{\sigma}{1-\kappa \sigma \tau_{1}}\right)
\sqrt{\tau_{1}^{}}.
\end{equation}
The coefficient of performance (COP), 
$\epsilon = Q_c/W$, is given by
\begin{equation}
    \epsilon = \frac{\kappa \sigma \theta}
    {1-\kappa \sigma \theta},
    \label{cop}
\end{equation}
which is fixed, for a given configuration 
of the refrigerator. 
The condition for Carnot bound $\epsilon < \epsilon_{\rm C}^{} = \theta/(1-\theta)$
requires $\kappa \sigma < 1$. 
The cooling power, $\dot{Q}_{c} = Q_c/t$,
is given by
\begin{equation}
\dot{Q}_{c}\left(\tau_{1}^{}\right)={K_{c} T_{c} \sigma}{\left(\frac{1}{\tau_{1}-1}+\frac{\sigma}{1-\kappa \sigma \tau_{1}^{}}\right)}^{-1}.
\end{equation}
For optimization of performance of the 
refrigerator, we choose 
a trade-off measure 
\cite{Yan1990}
between the cooling power and COP,
defined as $\chi = \epsilon \dot{Q}_{c}^*$.
For given values of parameters, 
the $\chi$-function is maximized
with respect to $\tau_{1}^{}$ 
at the optimal value given by 
Eq. (\ref{tau1st}).
The corresponding optimal $\chi$ value reads
\begin{equation}
{\chi}_{}^*=
K_{c}T_{c} \frac{(1- \kappa \sigma)\kappa \sigma \theta}{(1+\sqrt{\kappa})^{2}(1-\kappa \sigma \theta)}.
\end{equation}
Note that although $\tau_{1}^{*}$ has the same form
for both engine and refrigerator modes, the domain 
of parameter values must satisfy $\kappa \sigma <1$ here. We also notice the tradeoff
between the cooling power and 
the corresponding COP. As $\kappa \sigma \to 1$,
COP approaches the Carnot value. However, 
the cooling power vanishes, and so 
$\chi^*$ vanishes too.

So far, we see a close analogy between
the case of engine as treated 
in previous section and the refrigerator above. 
The efficiency and COP are both fixed 
for given parameters. 
As for the engine, we would wish to  
optimize the $\chi$-function over the 
parameter space.

Another optimization over the parameter $\sigma$,
yielding an optimal value:
    $\sigma^* = (1-\sqrt{1-\theta})/
    {k\theta}$, whereby
the COP at this point is 
\begin{equation}
   \epsilon_{\chi^*}^{} = \frac{1}{\sqrt{1-\theta}}-1 = \sqrt{\epsilon_{\rm C}^{}+1}-1. 
\end{equation}

At the optimal ratio $\sigma^{*}$, the ratios of cold to hot contact times is given by 
${1}/{\sqrt{\kappa}}$. Further, we
can calculate the explicit expressions
for each duration, along with 
other quantities, at optimal $\chi^*$.

\noindent 

\appendix

%\bibliography{mybib}
\end{document}